\documentclass[11pt]{AIAA}

\usepackage{subfig,float,hyperref}
\usepackage[usenames,dvipsnames]{color} 
\usepackage{amsfonts,amsmath}

\begin{document}

\title{Quantifying hazards: asteroid disruption in lunar distant retrograde orbits}

\author{Javier Roa\footnote{PhD Candidate, Applied Physics Department. Student Member AIAA\\ \emph{Present address:} Jet Propulsion Laboratory, California Institute of Technology, Pasadena, CA 91109, USA}}
\affiliation{Space Dynamics Group, Technical University of Madrid, Madrid, E-28040, Spain}
\author{Casey Handmer\footnote{PhD Candidate, Theoretical Astrophysics Including Relativity (TAPIR). Member APS, GGR.}}
\affiliation{Theoretical Astrophysics, California Institute of Technology, Pasadena, CA 91125, USA}

\begin{abstract}
	The Asteroid Redirect Mission (ARM) proposes to retrieve a near-Earth asteroid and position it in a lunar distant retrograde orbit (DRO) for later study, crewed exploration, and ultimately resource exploitation. During the Caltech Space Challenge, a recent workshop to design a crewed mission to a captured asteroid in a DRO, it became apparent that the asteroid's low escape velocity $(<1\,\mathrm{cm\,s}^{-1})$ would permit the escape of asteroid particles during any meaningful interaction with astronauts or robotic probes. This Note finds that up to 5\% of escaped asteroid fragments will cross Earth-geosynchronous orbits and estimates the risk to satellites from particle escapes or complete disruption of a loosely bound rubble pile.
\end{abstract}

\maketitle

\section{Introduction}

The Asteroid Redirect Mission (ARM) was the result of an initial feasibility study in 2012 by the Keck Institute for Space Studies~\cite{Keck}, proposing to retrieve a near-Earth asteroid and position it in a stable lunar orbit for later study, crewed exploration, and ultimately resource exploitation. The ARM is a bold concept, combining both emerging technologies and science, ranging from planetary defense to solar system formation to In-Situ Resource Utilization for crewed deep space missions. Requirements derived from expected $\Delta V$ capability and in-orbit stability dictate that a captured asteroid be placed in a Lunar distant retrograde orbit (DRO), or possibly Lunar Circulating Eccentric Orbits \cite{strange2013overview}.

DROs were originally found by H\'enon \cite{henon1969numerical}, who referred to them as the $f$ family, and typically encircle the $L_1$ and $L_2$ libration points. Proposed mission concepts exploiting DROs have focused on the Earth-Moon \cite{ming2009exploration} and Jupiter-Europa \cite{lam2005exploration} systems. The family of periodic orbits obtained in the circular restricted three body problem (CR3BP) is stable, but the stability region is immersed in a region of very unstable motion \cite{lara2007classification}. Some DROs around the Moon have been found to be stable for over 100 years, with the Sun's gravitational attraction the major external perturbation acting on the system \cite{bezrouk2014long}. The particular dynamics of the CR3BP has motivated recent studies on transfers to lunar DROs using ballistic capture manifold theory \cite{belbruno1993sun,parker2015low}. 

During the Caltech Space Challenge, a recent workshop to design a crewed mission to a captured asteroid in a DRO, it became apparent that the low escape velocity (\(<1\, \mathrm{cm\,s}^{-1}\)) would permit the escape of asteroid particles during any meaningful interaction with astronauts or robotic probes. This Note addresses the likely fate of escaped asteroid fragments and estimates the risk to Earth-orbiting satellites in the event of complete disruption of a loosely bound rubble pile.

\section{Stability of perturbed DROs}

In general, given a certain timescale, DROs possess a region of stability in phase space. Beyond this region, a perturbed particle will enter a resonance with the Moon that eventually (months to years) destabilizes the orbit. Subsequent motion is chaotic. Particles that escape the DRO enter orbits in cis-Lunar space for a period of time until they either collide with the Moon, or are ejected from the system entirely, as analysed in Section~\ref{stats}.

\begin{figure}[H]
  \centering
  \includegraphics[width=0.8\linewidth]{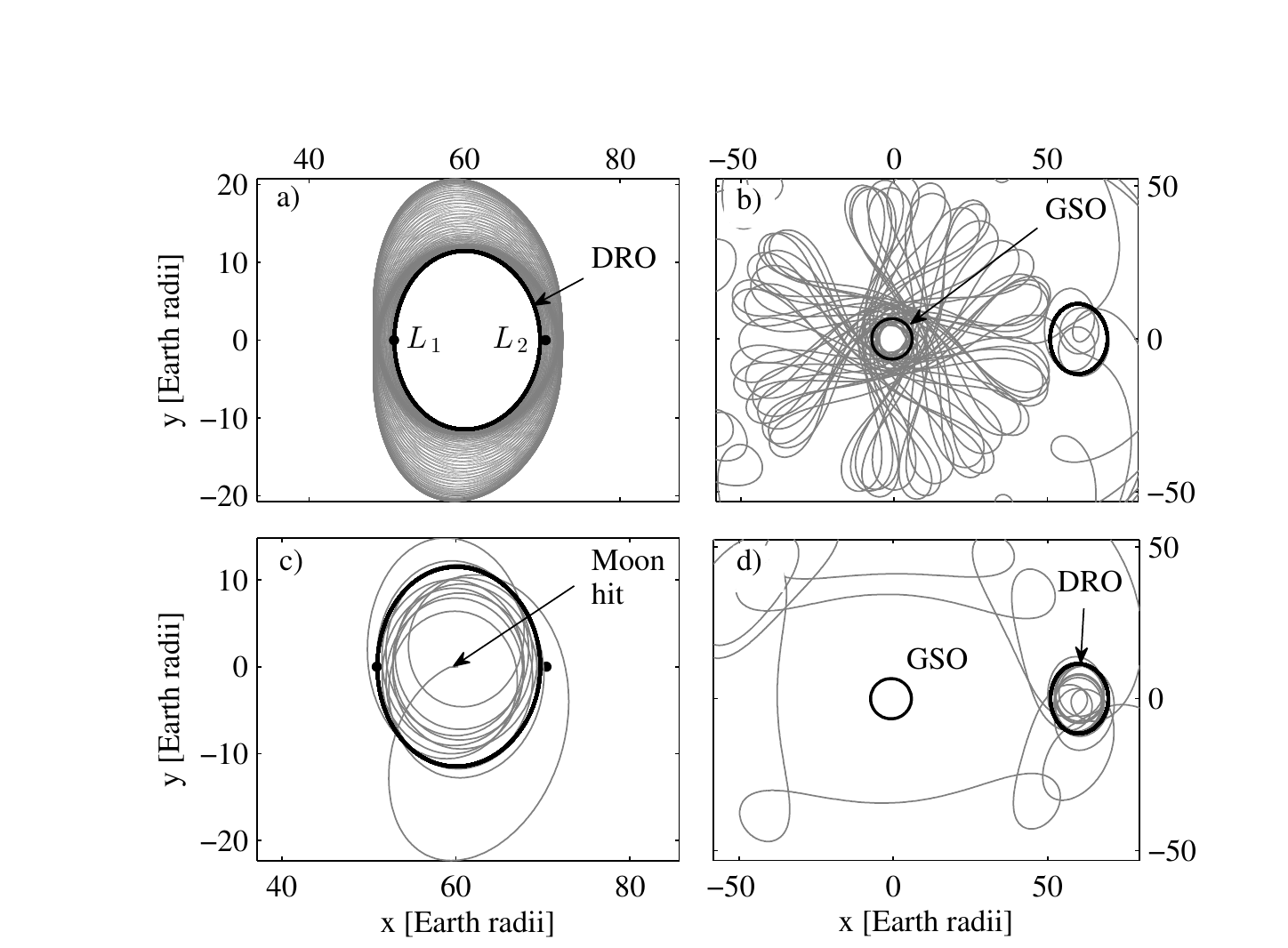}
\caption{\small Example of orbits in the CR3BP: a) Stable orbit, b) Resonant GSO crossings, c) Moon impact, d) Escape without crossing GSO nor impacting the Moon.\label{Fig:orbits}}  
\end{figure}

Fig.~\ref{Fig:orbits} shows examples of orbits in the CR3BP over a ten year period. A nominal DRO $(C=2.95)$ is perturbed by applying in-plane $\Delta V$s when crossing the $y$-axis close to $L_1$. Four cases are distinguished depending on the trajectory of the perturbed particles: a) the DRO remains stable over the considered time span, b) the perigee falls below GSOs, c) the particle impacts the Moon without crossing GSO, d) the particle escapes the Earth-Moon system without crossing GSO. Note that Lunar flybys are required to enter cis-Lunar regimes.

\begin{figure}[H]
  \centering
  \includegraphics[width=.7\textwidth]{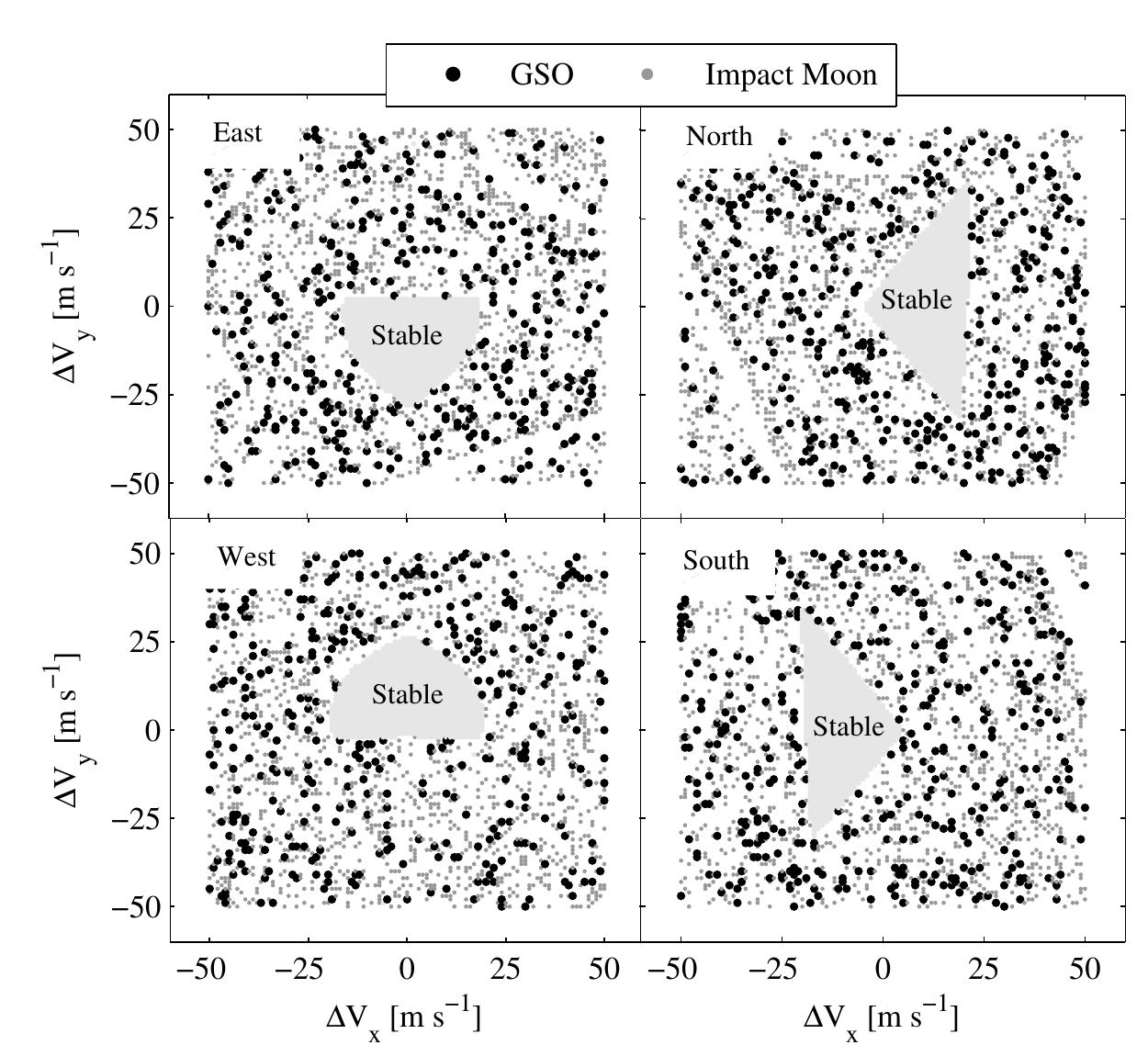}
  \caption{\small{Stability maps and fate of particles at the DRO's cardinal points. A region of stability is \(\mathcal{O}(10\,\mathrm{m\,s}^{-1})\) across. }}
  \label{fig:stabilityregions}
\end{figure}

Fig.~\ref{fig:stabilityregions} shows the stability regions at the DRO's four cardinal points generated by perturbing $\Delta V$s on a uniform, symmetric grid. Results of more general DROs were analysed statistically and presented in Table~\ref{Tab:distrib_DROs}. Qualitatively, the size of the stability region is similar, but the orientation and shape varies with orbital phase. The stability region grows with the Jacobi constant. In addition, for smaller values of $C$ the differences in the shape of the stability region across the cardinal points are more significant.

\section{Quantifying the risk}
\label{stats}

Table~\ref{table} studies the effect of the magnitude of the perturbation on the fate of the particles, showing the results from a campaign of Monte Carlo simulations. The higher the $\Delta V$, the more particles are ejected from the DRO through a Lunar flyby, increasing the number of GSO crossings and Moon impacts.

\begin{table}[H]
\centering
\begin{tabular*}{\linewidth}{@{\extracolsep{\fill}}l c c c@{} }
  \hline \hline
  Population &  \(<10\,\mathrm{m\,s^{-1}}\) & \(<25\,\mathrm{m\,s^{-1}}\) & \(<44\,\mathrm{m\,s^{-1}}\) \\
  \hline 
  Population size & 830    & 5404    & 10000  \\
  Escapes         & 10.1\% & 15.8\%  & 22.9\% \\
  GSO             & 2.2\%  & 3.4\%   & 5.3\%  \\
  Moon impact     & 10.7\% & 19.9\%  & 30.2\% \\
  Earth impact    & 0.0\%  & 0.0\%   & 0.0\%  \\
  Stable          & 77.0\% & 60.9\%  & 41.6\% \\
\hline\hline
\end{tabular*}
\caption{\small{Percentages of particle fates grouped by maximum perturbing \(\Delta V\) on a DRO defined by $C=2.95$. While smaller \(\Delta V\)s result in fewer GSO crossing particles, the cutoff is smaller than 5 m\,s$^{-1}$.}}
\label{table}
\end{table}

Table~\ref{Tab:distrib_DROs} presents the statistical results obtained for simulations based on different DROs, parameterized by the Jacobi constant $C$. Results in the table correspond to the mean value across the four cardinal points. We found that the results were robust with respect to applying the perturbation at different phases of the orbit, where statistical differences below $\pm0.5\%$ were observed. This is due to the non-chaotic nature of initially perturbed particles. While orbits with a higher Jacobi constant are more stable, achieving captured asteroid insertion requires unattainable $\Delta V$. Persistence time is also important; some of the candidate particles spent months within GSO over a 10 year period, vastly increasing their interaction cross section compared to transient meteor showers or interplanetary dust.

\begin{table}[H]
	\centering
	\begin{tabular*}{\linewidth}{@{\extracolsep{\fill}}lccccccc@{}}
		\hline\hline
		$C$ 	& $\langle r \rangle$ [km] & Period [days] &	GSO-crossing & Moon impact & $\dot{y}_\mathrm{f}\,[\mathrm{m\,s}^{-1}]$ & $\langle N_\mathrm{c}\rangle$ & $N_\mathrm{max}$ \\
		\hline 
		2.86 & 156898 & 21.59 & 6.81\% & 5.23\% & -663.93 & 41 & 864 \\
		2.89 & 124372& 18.86 & 4.99\% & 5.10\% & -590.30 & 49 & 906 \\
		2.92 & 91349& 14.98& 4.02\% & 6.40\% & -524.67 &  62 & 930  \\
		2.95 & 65810& 11.10&  5.10\% & 13.74\% & -485.44& 66 & 859\\
		2.98 & 49458& 8.21&  2.32\% & 7.70\% & -473.21 & 67 & 888 \\
		3.01 & 39149& 6.27 &  0.96\% & 5.18\% & -477.58 & 76 & 710 \\
		3.04 & 32214& 4.94 &  0.18\% & 1.70\% & -490.29 & 83 & 876 \\
		3.07 & 27290& 4.00 &  0.00\% & 0.00\% & -507.87 & $-$ & $-$\\
		3.10 & 23600& 3.30 &  0.00\% & 0.00\% & -527.81 & $-$ & $-$\\
		\hline\hline
	\end{tabular*}
	\caption{\small{Distribution of particle fates for different lunar DROs and $(\Delta V_x,\Delta V_y)\in[-50,50]\times[-50,50]\,\mathrm{m\,s^{-1}}$. Each orbit is defined by its mean radius, $\langle r \rangle$, period, and velocity at the far side of the Moon, $\dot{y}_\mathrm{f}$. Resonant returns are characterized by the average, $\langle N_\mathrm{c}\rangle$ and overall maximum, $N_\mathrm{max}$, of times that particles intersect GSO. 
\label{Tab:distrib_DROs}}}
\end{table}

\begin{figure}[H]
	\centering
	\includegraphics[width=0.8\textwidth]{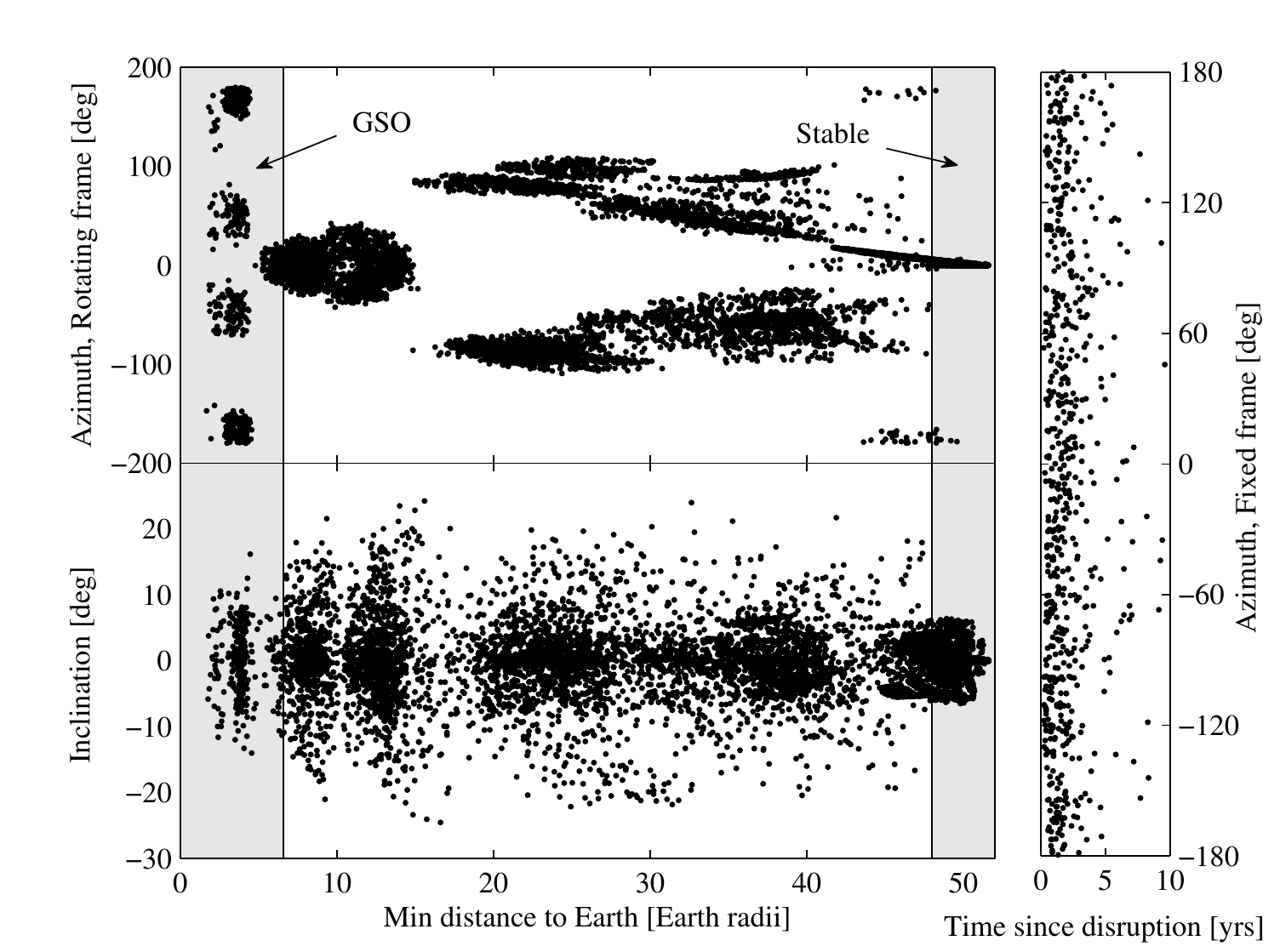}
	\caption{\small{Relative orientation of particles at closest approach for $(\Delta V_x,\Delta V_y)\in[-50,50]\times[-50,50]\,\mathrm{m\,s^{-1}}$ (planar case) and $(\Delta V_y,\Delta V_z)\in[-50,50]\times[0,100]\,\mathrm{m\,s^{-1}}$ (out-of-plane motion).\label{Fig:distribution_angles}}}
\end{figure}

Simulations show that the particles that get to GSO distance from the Earth do not reach low-Earth orbit. Risk analyses shall focus on the potential impact of asteroid fragments with satellites in GSO. The ARM preliminary mission design considers the early 2020s as the baseline for arriving to the lunar DRO. The inclination of the Moon's orbit with respect to the Earth's equator determines the incoming inclination of the potential impactors, and therefore the satellite orbits most likely to be affected. The relative inclination of the Moon reaches a minimum value of $18.2^\circ$ in Apr-2015, and a maximum of $29.0^\circ$ by Oct-2024. Close approaches detected in the Earth-Moon planar system may only interfere with GSO orbits at node crossings. Figure~\ref{Fig:distribution_angles} first studies the azimuthal distribution of particles at closest approach in the rotating reference. For particles entering GSO there exist four differentiated clusters, centered at $0^\circ$, $\pm45^\circ$ and $180^\circ$. We found that the mean time of flight to closest approach is roughly 23 months, with a standard deviation of 20 months. The orientation of the incoming flux in the fixed frame shows that an isotropic distribution of particles may be expected in the first four years. Such a distribution suggests that particles reaching GSO distance from Earth may eventually cross GEO at the ascending/descending node. 

Fig.~\ref{Fig:distribution_angles} also presents the inclinations of incoming particles at closest approach relative to the orbital plane of the Moon, when applying out-of-plane $\Delta V_z$ of up to $100\,\mathrm{m\,s^{-1}}$. The mean inclination is $0.2^\circ$, with a standard deviation of $5.1^\circ$. This translates into an effective radius of the cross section of the particle flux of 3760 km ($1\sigma$). Hence, 5.7\% of GEO will be covered by the incoming flux, considering both the ascending and descending nodes and the flux cross section at $1\sigma$. If $\Delta V_z\leq 50\,\mathrm{m\,s^{-1}}$ the standard deviation reduces to $3.3^\circ$, and the coverage on GEO is 3.6\% $(1\sigma)$.

In the event of complete asteroid disruption into small particles, an ARM design reference \(500\,\mathrm{T}\) asteroid would put \(25\,\mathrm{T}\) of particles inside GSOs, corresponding to a flux of \(34\,\mathrm{kg\,s}^{-1}\). In contrast to resonant DRO escapees, typical interplanetary cosmic dust passes through the GSO region in around 30 minutes, with a baseline flux of around \(100\,\mathrm{kg\,s}^{-1}\)\cite{ref:dust}. {\it Ipso facto} the catastrophic disruption of a single 5 m, 500 T asteroid would increase dust risk to satellites by more than 30\%. With a mean persistence time of around a decade, a flux of only 500\,kg/day of particles from asteroids in DROs will more than {\it double} the risk to satellites, necessitating the establishment of good practices for asteroid exploration and exploitation in cis-Lunar space.

\section{Conclusions}
On average we found that $5\%$ of the particles escaping from a lunar DRO intersect geosynchronous altitude, with no immediate risk to low-Earth orbits. The incoming flux has been found to be effectively isotropic and might affect satellites in geostationary orbit (GEO) at node crossings. In 10 yrs particles have been found to intersect GEO up to 900 times, 63 times on average. A uniform flux might be expected during a 40 months period $(1\sigma)$. Out-of-plane perturbations yield flux cross sections which cover 6\% $(1\sigma)$ of GEO. These results might serve as a statistical basis for detailed analyses of collision risk in particular cases, together with analytical studies on the propagation of fragment clouds \cite{letizia2015analytical}. 

\section*{Acknowledgments}
This research was motivated by the Caltech Space Challenge 2015. Authors thank Team Voyager, the organizers, and the sponsors: Northrop Grumman, Lockheed Martin, KISS, JPL, SpaceX, Millenium Space Systems, Galcit and the Moore-Hufstedler Fund for Student Life at Caltech. JR thanks ``La Caixa'' for his doctoral fellowship. CH thanks Damon Landau for thought provoking discussions.

\end{document}